\documentstyle[multicol,prl,aps,epsfig]{revtex}
\renewcommand{\narrowtext}{\begin{multicols}{2}
\global\columnwidth20.5pc\noindent}
\renewcommand{\widetext}{\end{multicols}
\global\columnwidth42.5pc}
\begin{document}
\draft
\preprint{OUCMT-99-6}
\title{Middle-Field Cusp Singularities in the Magnetization Process of One-Dimensional Quantum Antiferromagnets}
\author{Kouichi Okunishi, Yasuhiro Hieida and Yasuhiro Akutsu}
\address{Department of Physics, Graduate School of Science, Osaka University,\\
Machikaneyama-cho 1-1, Toyonaka, Osaka 560-0043, Japan.
}
\date{\today}
\maketitle

\begin{abstract}
We study the zero-temperature magnetization process ($M-H$ curve) of one-dimensional quantum antiferromagnets using a variant of the density-matrix renormalization group method.
For both the $S=1/2$ zig-zag spin ladder and the $S=1$ bilinear-biquadratic chain, we find clear cusp-type singularities in the middle-field region of the $M-H$ curve.
These singularities are successfully explained in terms of the double-minimum shape of the energy dispersion of the low-lying excitations.
For the $S=1/2$ zig-zag spin ladder, we find that the cusp formation accompanies the Fermi-liquid to non-Fermi-liquid transition.

\end{abstract}

\narrowtext

Low-dimensional antiferromagnetic (AF) quantum spin systems with various spin magnitude $S$ and various spatial structures, have been a field of active researches, both experimentally and theoretically.
In particular, the magnetization process ($M-H$ curve, $M$: magnetization, $H$:  magnetic field) of AF spin chain at low-temperatures have recently drawn much attention, because it exhibits various phase-transition-like behaviors:
e.g., the critical phenomena $\Delta M\sim \sqrt{H-H_{\rm c}}$ associated with the gapped  excitation  (excitation gap $\propto H_{\rm c}$)\cite{Aff1,SaTa1,OHA} or with the saturated magnetization (at the saturation field $H_{\rm s}$),\cite{SaTa1,OHA,PaBo,HoP,JF,KA} magnetization plateau,\cite{plateau} and, the first-order transition.\cite{Sato-Akutsu}
These are field-induced phase transitions of the ground state, which reflect non-trivial energy-level structure at zero field.

What we consider in this Letter is another type of singularity which has not been discussed so much: the cusp singularity at $H=H_{\rm cusp}$ in the middle-field region ($H_{\rm c}<H_{\rm cusp}<H_{\rm s}$). 
Existence this type of singularity was first demonstrated by Parkinson\cite{Parkinson} for the integrable Uimin-Lai-Sutherland (or, SU(3)) chain,\cite{UiLaiSut} and has also been known for some other integrable spin chains,\cite{SU(N),Hubbard,SpinAlt,SU3-D} and ladders. \cite{IntegLadder3,IntegLadder1}
Derivation of the magnetization cusp for each model, however, relies on the model's integrability in an essential manner, restricting the Hamiltonian to be of somewhat unrealistic form.
Hence, whether or not this type of singularity can be found in realistic systems is a highly non-trivial question.  In the present Letter, we make the first systematic numerical study of the ``middle-field cusp singularity'' (MFCS, for short) in the zero-temperature $M-H$ curve for non-integrable systems, through which we give a positive answer for the above question.
Namely, we show that the $S=1/2$ zig-zag spin ladder and the $S=1$ bilinear-biquadratic chain actually exhibit MFCS in the $M-H$ curve.

The numerical method we employ is the product-wavefunction renormalization group (PWFRG),\cite{PWFRG}  which is a variant of the density-matrix renormalization group (DMRG).\cite{DMRG}
Efficiency of the PWFRG in calculation of the $M-H$ curve has been demonstrated in previous studies;\cite{OHA,HOA}
the PWFRG will allow us to obtain the $M-H$ curve {\em in the thermodynamic limit} with enough accuracy to detect ``weak'' (non-divergent) singularity like the MFCS.

Consider the $S=1/2$ zig-zag spin ladder, whose actual realization can be made as a quasi-one-dimensional material.\cite{LadderExp} The $M-H$ curve of the system with bond alternation has recently been studied, where the cusp singularity has not been discussed.\cite{zig-zag}
The Hamiltonian of the system is given by
\begin{eqnarray}
{\cal H}_{\rm zig-zag}&=&\sum_{i} [\vec{S}_{i}\cdot\vec{S}_{i+1}
+J  \vec{S}_{i}\cdot\vec{S}_{i+2}] 
- H \sum_i S_i^z, \label{ladder}
\end{eqnarray}
where $\vec{S}_i$ is the $S=1/2$ spin operator at the $i$-th site.  We have normalized the nearest neighbor coupling to unity, and have denoted the next-nearest coupling by $J$ ($>0$).
In Fig. 1, we show the $M-H$ curve calculated by using the PWFRG.  We see the $M-H$ curve for  $J>1/4$ exhibits a clear MFCS.


We take down-spin-particle picture, where one down spin in the saturated (all up) state is regarded as a particle.  Then the system near $H_s$ can be regarded as that of interacting Bose particles, which reduces to the well-known delta-function Bose gas ($\delta$-BG) model\cite{LiLi} in the low-energy limit.\cite{SaTa1,OHA,JF}
Important point is that the $\delta$-BG at low density is equivalent to spinless free-Fermi gas.\cite{OHA,HoP,LiLi}
Hence, for discussion of the $M-H$ curve near $H_{s}$, we may treat the system as that of the spinless Fermions which is completely characterized by the one-particle excitation energy dispersion $\omega(k)$.

 The one-down-spin excitation energy $\omega(k)$ is calculated to be 
\begin{equation}
\omega(k)= \cos k-1 +J(\cos 2k-1), 
\label{zzdisp}
\end{equation}
which we depicted in Fig.2. 
It should be noted that, at $J=1/4$, there occurs a qualitative change in the shape of $\omega(k)$:
For $J\le 1/4$, $\omega(k)$ have a single minimum at $k=\pi$, while, for $J>1/4$, $\omega(k)$ at $k=\pi$ changes into a local maximum and two minima newly appear (corresponding $k$-positions are determined from $\cos k=-1/(4J)$).\cite{IntegLadder3,saturation}
Then, the van Hove singularity associated with the double-minimum shape of $\omega(k)$ gives a simple explanation of the MFCS in the $M-H$ curve for $J>1/4$.\cite{IntegLadder1}


Let us make a quantitative analysis as follows.
From the bottom of $\omega(k)$, we obtain
$H_s= 2$ for $1/4 \ge J \ge 0$, and $H_s=1 + 2J + 1/(8J)$ for $ J \ge 1/4$.
As the applied field $H$ is decreased below $H_{s}$, the down-spin density becomes to be non-zero. The $M-H$ curve is obtained from
\begin{eqnarray}
M&=&1/2-\frac{1}{2\pi}\int R(k) dk, \label{ffm} \\
E(M)&=&\frac{1}{2\pi}\int \omega(k)R(k) dk, \label{ffe} \\
H&=&\frac{\partial E(M)}{\partial M},\label{ffh}
\end{eqnarray}
where $R(k)$ is the zero-temperature Fermi distribution function which is unity inside the ``Fermi vacuum'' but is zero otherwise. Hence, how the particles fills the energy band tells us the essential behavior of the $M-H$ curve.

In $0\le J \le 1/4$, the $M-H$ curve is smooth in the whole range of $0\le M \le1/2$.
The one-particle energy $\omega(k)$ near the bottom position $k=\pi$ has the expansion $\omega(k)=-2+\frac{1}{2}(1-4J)(\Delta k)^2+\frac{1}{24}(16J-1)(\Delta k)^4+\ldots $ where we have introduced $\Delta k=k-\pi$.
The $H-M$ curve near $H_{s}$ is then calculated from (\ref{ffm}), (\ref{ffe}), and (\ref{ffh}) to be 
\begin{equation}
H_{s}-H= \frac{\pi^2(1-4J)}{2}(\Delta M)^2 + \frac{\pi^4(16J-1)}{24}(\Delta M)^4 \cdots ,
\end{equation}
where $\Delta M =1/2 -M$.  In the correctly mapped $\delta$-BG treatment for a class of AF spin chains,\cite{OHA}
there may be $(\Delta M)^3$ term due to the finiteness of the effective coupling.  Hence, for the present model, we fit the PWFRG-calculated $M-H$ curve with
\begin{equation}
H_{s}-H= \alpha (\Delta M)^2 \left[1+ \gamma (\Delta M)+\delta (\Delta M)^2 \dots \right], \label{fit2}
\end{equation}
to check whether the obtained value of $\alpha$ agrees with the free-Fermion prediction.  The best-fit results are $\alpha=2.6$ ($J=0.1$) and $\alpha=0.92$ ($J=0.2$), which are consistent with the free-Fermion prediction: $\alpha=2.960\ldots$ ($J=0.1$) and $\alpha=0.9869\ldots$ ($J=0.2$).

At $J=1/4$, $(\Delta k)^2$-term in $\omega(k)$ vanishes, leading to a different form of the expansion: $H_{s}-H= \tilde{\alpha}(\Delta M)^4\left[1+\tilde{\gamma} (\Delta M)+\tilde{\delta} (\Delta M)^2  \cdots \right] $.
The best-fit value of $\tilde{\alpha}$ from the PWFRG result lies in the range $\tilde{\alpha}=14\sim17$,\cite{error} which is in reasonable agreement with the free-Fermion prediction $\tilde{\alpha}=12.2\ldots$.
Hence the free-Fermion picture also holds for $J=1/4$, and we have $\Delta M\sim (H-H_s)^{1/4}$ with the critical exponent $1/4$ being different from the ``standard'' value $1/2$,\cite{OHA,PaBo,HoP,KA} supporting the finite-size diagonalization result.\cite{unusual}

We have thus seen that the system for $J\leq 1/4$ is a Fermi liquid, near $H_{s}$.
Hence, also for $J>1/4$, we may expect that the system continues to behave as a Fermi liquid.  Qualitatively, the Fermi-liquid character well explains the $M-H$ curve, in particular, the appearance of the MFCS.
For quantitative discussion, however, there emerges an important difference from the case of $J\le1/4$:
Due to the double-minimum shape of $\omega(k)$, the system becomes to be {\em two-component} liquid [each component is composed of modes around each minimum].
Since the separation into two components may not be complete, there should remain interactions between the components.  Such a correlated multicomponent system may often behave as a non-Fermi liquid, or, Tomonaga-Luttinger (TL, for short) liquid,\cite{TL,Solyom,Haldane} whose typical example is the Hubbard chain.\cite{HubbardTL}

The TL liquid is characterized by the smooth edge of the momentum distribution at the Fermi point $k_{\rm F}$: $R(k)\sim (k_{\rm F}-k)^{\zeta}$ ($k\sim k_{\rm F}$, $\zeta>0$ ).\cite{Haldane}
Then, for very small particle density, we assume
\begin{equation}
R(k)=R_{0}(k_{\rm F} -|k|)^{\zeta}/k_{\rm F}^{\zeta},
\end{equation}
where $R_{0}$ is a constant, and we have assumed that $R(0)$ remains to be finite even in the vanishing particle density ($k_{\rm F}\rightarrow0$).
Assuming that each component has a parabolic energy dispersion $\omega(k)=\sigma \Delta k^2$ ($\Delta k=k-k^{*}$) around each minimum $k=k^{*}$, we have the following form of the ground-state energy density $E_{\rm G}$ as a function of the total particle number density $\rho$:
\begin{eqnarray}
E_{\rm G}&=&C_{\rm TL}\frac{\sigma\pi^2}{12}\rho^3, \label{EG-rho}\\
        C_{\rm TL}&=&C_{\rm TL}(R_{0},\zeta)\equiv \frac{24(\zeta+1)^2}{R_{0}^2(\zeta+2)(\zeta+3)}.\label{nonFFfactor}
\end{eqnarray}
For the two-component (non-interacting) Fermi liquid, we have $\zeta=1$ and $R_{0}=2$ giving $C_{\rm TL}(R_{0},0)=1$.
Therefore, deviation of $C_{\rm TL}$ from unity implies the non-Fermi liquid character of the system.

Since $\rho$ corresponds to $\Delta M=1/2-M$, the TL-liquid expression (\ref{EG-rho}) gives the coefficient $\alpha$ in (\ref{fit2}) as $\alpha=C_{\rm TL}\sigma\pi^2/4$ where $\sigma$ is calculated from the band curvature at each minimum of $\omega(k)$.  Explicitly, we have $\sigma=(16J^2-1)/(8J)$ and
\begin{equation}
\alpha=C_{\rm TL} \frac{(16J^2-1)}{32J}\pi^2. \label{alphaTL}
\end{equation}
The best-fit values of $\alpha$ from the PWFRG calculation are $5.9\pm0.3$ ($J=0.4$) and $7.4\pm0.4$ ($J=0.5$).\cite{error}
These values {\em disagree} with the free-Fermion values ((\ref{alphaTL}) with $C_{\rm TL}=1$): $\alpha=1.20\ldots$ ($J=0.4$) and $\alpha=1.85\ldots$ ($J=0.5$).  Rather, our calculation leads to $C_{\rm TL}\approx 4$ in (\ref{alphaTL}), showing a clear sign of the non-Fermi-liquid character of the system.
Remarkably, in the Hubbard chain with on-site interaction $U$ ($>0$) which can be viewed as a two-chain $S=1/2$ spin ladder via the Jordan-Wigner transformation, the same factor $C_{\rm TL}=4$ appears in the $U=\infty$ limit\cite{Shiba} or in the low-density limit.\cite{UKO}  Therefore, we may conclude that, near $H_{s}$, the $S=1/2$ zig-zag ladder for $J>1/4$ is a two-component non-Fermi TL liquid. 

Let us next consider the bilinear-biquadratic (BLBQ, for short) chain with the Hamiltonian
\begin{eqnarray}
{\cal H}_{\rm BLBQ} &=& \sum_{i}\left[\vec{S}_{i}\cdot\vec{S}_{i+1} + {\beta}(\vec{S}_{i}\cdot\vec{S}_{i+1})^2\right]  - H \sum_{i}S_{i}^{z},
\label{BLBQ}
\end{eqnarray}
where $\vec{S}_{i}=(S_{i}^{x},S_{i}^{y},S_{i}^{z})$ is the $S=1$ spin operator at the site $i$.\cite{BLBQchain}  In Fig.3 we show PWFRG-calculated $M$-$H$ curves for $\beta=0.45$, $0.6$, $0.8$, $1.0$, whare clear MFCS is seen.\cite{comment}


  In Ref.\cite{OHA}, we made a quantitative test for the square-root critical behavior,\cite{Aff1,SaTa1,SaTa-new} $M\sim A(H-H_{c})^{1/2}$ ($A$: amplitude), of the $M-H$ curve near the lower critical field $H_{c}$ (which is proportional to the excitation gap at $H=0$).
There, we found a critical value $\beta_{c}$ ($\sim0.41$) where the critical exponent changes from 1/2 to 1/4, which is in close similarity to the $J=1/4$ case of the $S=1/2$ zig-zag spin ladder.
We should remark that the change in the shape of $\omega(k)$ at $\beta_{c}$ has recently been found.\cite{disp}  Therefore, the MFCS in the BLBQ chain is well explained in terms of the Fermi/Tomonaga-Luttinger-liquid picture.
We should also note that the observed two-component Fermi/Tomonaga-Luttinger-liquid behavior for $\beta>\beta_{c}$,\cite{two-comp-TL} is also consistent with the appearance of the MFCS.

To summarize, in this Letter we have studied the middle-field cusp singularity (MFCS) in the zero-temperature magnetization process ($M-H$ curve) for antiferromagnetic spin systems in one dimension.
For the $S=1/2$ zig-zag spin ladder and the  $S=1$ bilinear-biquadratic chain, we have found clear MFCS in the $M-H$ curve obtained by using the product-wavefunction renormalization group method which is a variant of the density-matrix renormalization group (DMRG) method.
We have explained the mechanism for the MFCS in terms of the shape-change in the energy dispersion curve of the low-lying excitation.
Further, for the  $S=1/2$ zig-zag spin ladder, we have shown that the formation of the MFCS accompanies the Fermi-liquid to non-Fermi-liquid (Tomonaga-Luttinger-liquid) transition in the character of the system.

As far as we know, what we have found for the $S=1/2$ zig-zag spin ladder is the first non-trivial example of physically observable MFCS.\cite{IntegLadder2}  For actual experimental observation, we should of course take the finite-temperature effect into account, which can be made by the ``finite-temperature DMRG''.\cite{finiteT}
As an important implication drawn from the present study, we should note that the essential mechanism for appearance of the MFCS is the multi-minimum structure of the low-lying excitation energy.
Such structure may be well expected for systems with non-trivial spatial structures and/or competing interactions, which often accompany incommensurability in physical quantities.
In fact, for the  $S=1/2$ zig-zag spin ladder in the large $J$ region ($J>0.5$) (incommensurability at zero field is reported\cite{incomm}), we have observed another cusp near the {\em lower critical field}, whose details will be published elsewhere.

This work was partially supported by the Grant-in-Aid for Scientific Research from Ministry of Education, Science, Sports and Culture (No.09640462), and by the ``Research for the Future'' program of the Japan Society for the Promotion of Science (JSPS-RFTF97P00201). One of the authors (K. O.) is supported by JSPS fellowship for young scientists.

\begin{figure}
\epsfig{file=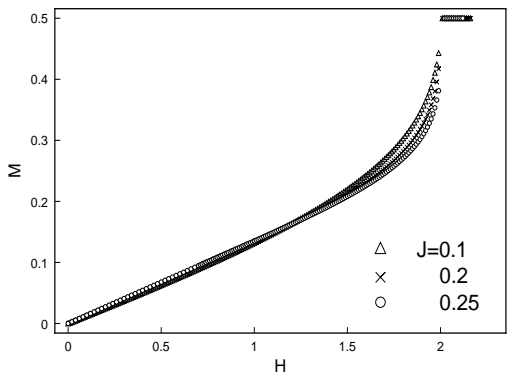,height=50mm,width=70mm}

\epsfig{file=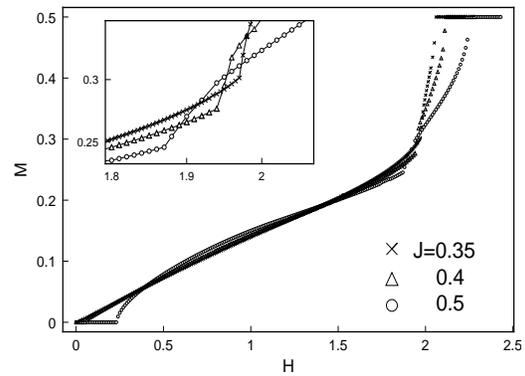,height=50mm,width=70mm}
\caption{The $M-H$ curve of the zig-zag spin chain calculated by the PWFRG with number of the  retained bases $m=30$. (a) $J=0.1$, $0.2$, and $0.25$. (b) $J=0.35$, $0.4$, and $0.5$. Inset: magnification of the curves around the cusps.}
\end{figure}

\begin{figure}
\epsfig{file=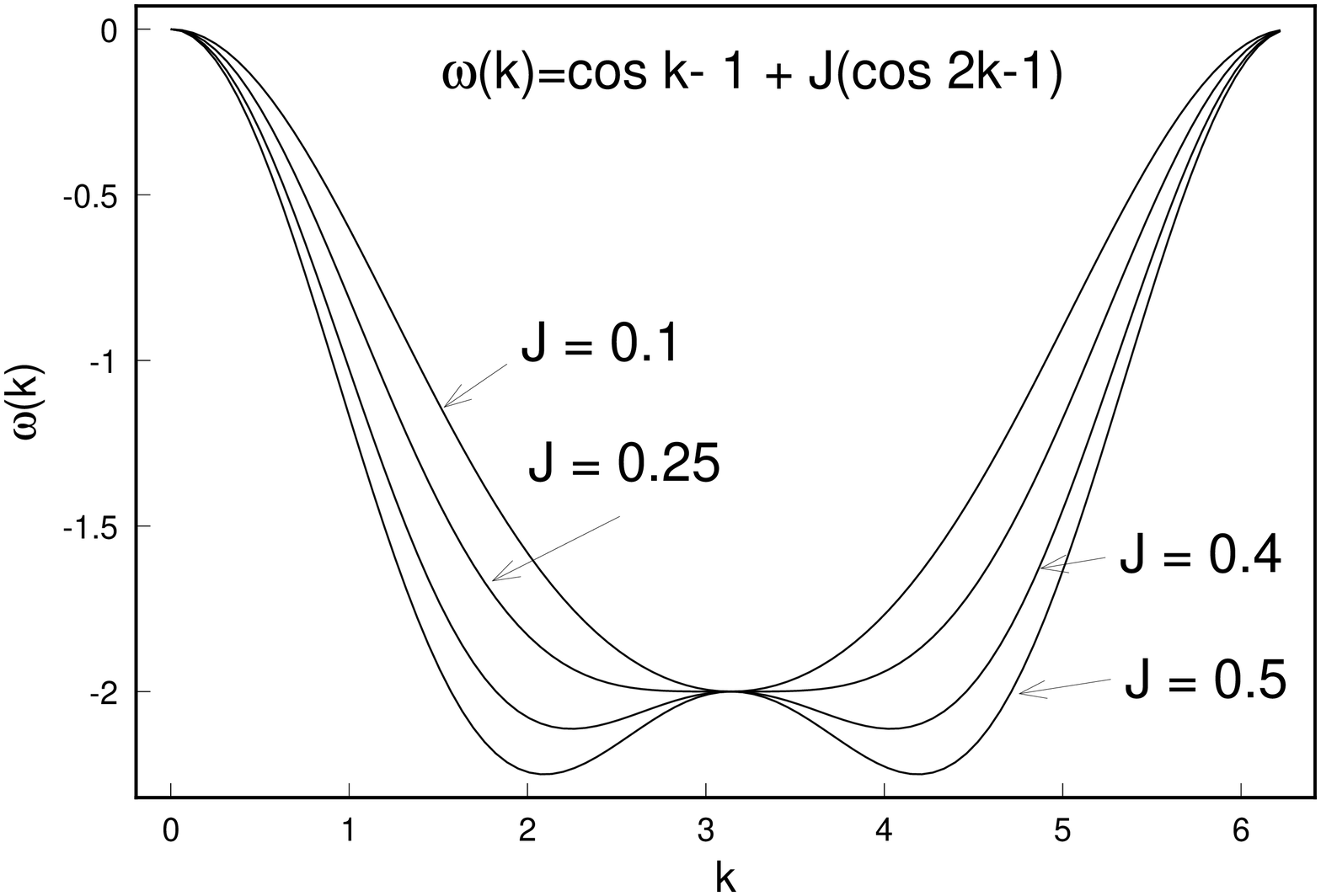,height=50mm,width=70mm}
\caption{The dispersion curve of the ``one-down-spin'' for the zig-zag spin chain.}
\end{figure}

\begin{figure}
\epsfig{file=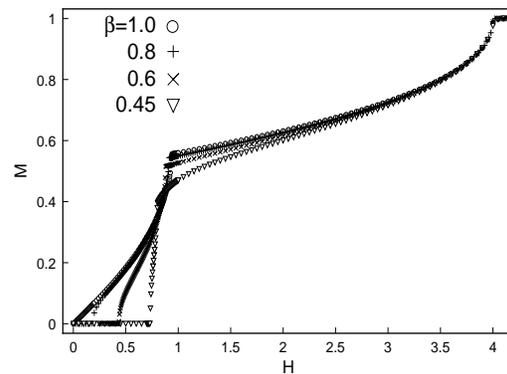,height=50mm,width=70mm}
\caption{The $M-H$ curve of the BLBQ chain for $\beta=0.45$, $0.6$, $0.8$ and $1.0$ calculated by the PWFRG with number of the  retained bases $m=70$.}
\end{figure}

\widetext

\begin{references}
\bibitem{Aff1} I. Affleck, Phys. Rev. B {\bf 43}, 3215 (1991);
E. S{\o}rensen and I. Affleck, Phys. Rev. Lett. {\bf 71}, 1633 (1996).

\bibitem{SaTa1} M. Takahashi and T. Sakai,  J. Phys. Soc. Jpn. {\bf 60}, 760 (1991).

\bibitem{OHA} K. Okunishi, Y. Hieida amd Y. Akutsu, Phys. Rev. B {\bf 59}, 6806 (1999).

\bibitem{PaBo} J.B. Parkinson and J.C. Bonner, Phys. Rev. B {\bf 32}, 4703 (1985).

\bibitem{HoP} R.P. Hodgeson and J.B. Parkinson. J. Phys. C {\bf 18}, 6385 (1985).

\bibitem{JF} M.D. Johnson and M. Fowler, Phys. Rev. B{\bf 34}, 1728 (1986).

\bibitem{KA} H. Kiwata and Y. Akutsu, J. Phys. Soc. Jpn. {\bf 63}, 3598 (1994).

\bibitem{plateau} T. Tonegawa, T. Nakao and M. Kaburagi, J. Phys. Soc. Jpn. {\bf 65}, 3317 (1996); K. Totsuka, Phys. Lett. A {\bf 228}, 103 (1996); M. Oshikawa, M. Yamanaka and I. Affleck, Phys. Rev. Lett. {\bf 78}, 1984 (1997).

\bibitem{Sato-Akutsu} R. Sato and Y. Akutsu, J. Phys. Soc. Jpn {\bf 65}, 1885 (1996).

\bibitem{Parkinson} J.B. Parkinson, J. Phys. Condens. Matter {\bf 1}, 6709 (1989).

\bibitem{UiLaiSut} G.V. Uimin, JEPT Lett. {\bf 12}, 225 (1970); C.K. Lai, J. Math. Phys. {\bf 15} (1974) 1675.; B. Sutherland,  Phys. Rev. B {\bf 12}, 3795 (1975).

\bibitem{SU(N)} H. Kiwata and Y. Akutsu, J. Phys. Soc. Jpn. {\bf 63}, 4269 (1994).

\bibitem{Hubbard} H. Kiwata, J. Phys. Condens. Matter {\bf 7}, 5045 (1995).

\bibitem{SpinAlt}  A.A. Zvyagin, and  P. Schlottmann, Phys. Rev. B {\bf 52}, 6569 (1995); M. Fujii, S. Fujimoto and N. Kawakami, J. Phys. Soc. Jpn. {\bf 65}, 2381 (1996).

\bibitem{SU3-D} A. Schmitt, K.- H. M{\"u}tter and M. Karbach, J. Phys. A {\bf 25}, 4721 (1996).

\bibitem{IntegLadder3}  V.Yu. Popkov and A.A. Zvyagin, Phys. Lett. {\bf A175}, 295 (1993); A.A. Zvyagin, Phys. Rev. B {\bf 57}, 1035 (1998).


\bibitem{IntegLadder1} H. Frahm and C. R{\"o}denbeck, J. Phys. A {\bf 30}, 4467 (1997).


\bibitem{PWFRG} T. Nishino and K. Okunishi, J. Phys. Soc. Jpn. {\bf 64}, 4084  
(1995).

\bibitem{DMRG} S.R. White, Phys. Rev. Lett. {\bf 69}, 2863 (1992); 
Phys. Rev. {\bf B 48}, 10345 (1993).

\bibitem{HOA} Y. Hieida, K. Okunishi and Y. Akutsu, Phys. Lett. A {\bf 233},  464 (1997).

\bibitem{LadderExp} M. Matsuda and K. Katsumata, J. Mag. Mag. Mat. {\bf 140-145}, 1671 (1995).


\bibitem{zig-zag} T. Tonegawa, T. Nishida and M. Kaburagi, Physica B {\bf 246}, 368 (1998); K. Totsuka, Phys. Rev. B {\bf 57}, 3454 (1998).

\bibitem{LiLi} E. Lieb and W. Liniger, Phys. Rev. {\bf 130}, 1605 (1963).

\bibitem{saturation}D.C. Carba, A. Honecker, P. Pujol, cond-mat//9902112.

\bibitem{error} The error-bar depends on the choice of the fitting window.

\bibitem{unusual} M. Schmidt, C. Gerhardt, K.-H. M{\"u}tter and M. Karbach, J. Phys. Condens. Matter. {\bf 8}, 553 (1996).

\bibitem{TL} S. Tomonaga, Prog. Theor. Phys. {\bf 5}, 544 (1950); J.M. Luttinger, J. Math. Phys. {\bf 4}, 1154 (1963); D.C. Mattis and E.H. Lieb, J. Math. Phys. {\bf 6}, 304 (1965). 

\bibitem{Solyom} J. Solyom, Adv. Phys. {\bf 28}, 201 (1979).

\bibitem{Haldane} F.D.M. Haldane, Phys. Rev. Lett. {\bf 45}, 1358 (1980); J. Phys. C {\bf 45}, 2585 (1981).

\bibitem{HubbardTL} H. Frahm and V.E. Korepin, Phys. Rev. B {\bf 42}, 10533 (1990); N. Kawakami and S.-K. Yang, Phys. Lett. A {\bf 148}, 359 (1990); H.J. Schulz, Phys. Rev. Lett. {\bf 64}, 2831 (1990).

\bibitem{Shiba} H. Shiba, Phys.Rev. B {\bf 6}, 930 (1972).
\bibitem{UKO} T. Usuki, N. Kawakami and A. Okiji, Phys. Lett. A {\bf 135}, 476 (1989).

\bibitem{BLBQchain} R.J. Bursill, T. Xiang and G.A. Gehring, J. Phys. Math. Gen. {\bf 28}, 2109 (1995); U. Schollw\"ock, Th. Jolic{\oe}re and T. Garel, Phys. Rev. B {\bf 53}, 3304 (1996).

\bibitem{comment} Existence of the MFCS for BLBQ chain was briefly commented in Ref.\cite{OHA}.

\bibitem{SaTa-new} T. Sakai and M. Takahashi, Phys. Rev. B {\bf 57}, R8091 (1998).
\bibitem{disp} O. Golinelli, Th. Jolic{\oe}ur and  E.S. S{\o}rensen, cond-mat/9812296.

\bibitem{two-comp-TL} G. F\'ath and B. Littlewood,  Phys. Rev. B {\bf 58}, R14709 (1998).

\bibitem{IntegLadder2} An argument based on the integrable model\cite{IntegLadder1} is recently given by H. Frahm and C. R{\"o}denbeck (cond-mat/9812103, to appear in, Eur. Phys. J. B).

\bibitem{finiteT} X. Wang and T. Xiang, Phys. Rev. B {\bf 56}, 2221 (1997); N. Shibata, J. Phys. Soc. Jpn. {\bf 66}, 5061 (1997); K. Okunishi, preprint.

\bibitem{incomm} S. Watanabe and H. Yokoyama, to appear in J. Phys. Soc. Jpn., and references cited therein.



\end{references}
\end{document}